\documentclass[twocolumn,twoside,aps,superscriptaddress,prl,showpacs]{revtex4}
\usepackage{graphicx}
\usepackage{dcolumn}
\usepackage{amssymb,amsmath}
\newcommand{\BB}{B}
\newcommand{\KK}{K}
\newcommand{\NN}{N}
\newcommand{\ee}{\nu}
\newcommand{\AAA}{A}
\begin{document}
\bibliographystyle{apsrev}

\title{Spontaneous Ratchet Effect in a Granular Gas}

\author{Devaraj van der Meer}
\affiliation{Department of Applied Physics and J.M. Burgers Centre
for Fluid Dynamics, University of Twente, P.O. Box 217, 7500 AE
Enschede, The Netherlands}
\author{Peter Reimann}
\affiliation{Department of Physics, University of Bielefeld, 33615
Bielefeld, Germany}
\author{Ko van der Weele}
\affiliation{Department of Applied Physics and J.M. Burgers Centre
for Fluid Dynamics, University of Twente, P.O. Box 217, 7500 AE
Enschede, The Netherlands}
\author{Detlef Lohse}
\affiliation{Department of Applied Physics and J.M. Burgers Centre
for Fluid Dynamics, University of Twente, P.O. Box 217, 7500 AE
Enschede, The Netherlands}


\begin{abstract}
The spontaneous clustering of a vibrofluidized granular gas is
employed to generate directed transport in two different
compartmentalized systems: a \emph{``granular fountain''} in which
the transport takes the form of convection rolls, and a
\emph{``granular ratchet''} with a spontaneous particle current
perpendicular to the direction of energy input. In both instances,
transport is not due to any system-intrinsic
anisotropy, but arises as a spontaneous collective
symmetry breaking effect of many interacting granular particles.
The experimental and numerical results are quantitatively
accounted for within a flux model.
\end{abstract}

\maketitle

The conversion of random fluctuations into directed motion in a
periodic system -- termed ratchet effect -- is of interest in a
wide variety of physical, biological, and technological contexts
\cite{Linke02}. Overall, two necessary conditions for the
emergence of the effect have been identified \cite{Reimann02},
namely (i) the absence of thermal equilibrium and (ii) the
breaking of the inversion symmetry. The most common realization of
the latter condition is by means of an intrinsic anisotropy in the
system, e.g. a spatially periodic but asymmetric potential, or a
so-called dynamical asymmetry due to a time-dependent driving
force with zero mean but non-vanishing higher moments
\cite{Reimann02}. In this Letter, we are concerned with {\em
perfectly symmetric systems} where the symmetry breaking arises
spontaneously as a collective effect of many interacting particles
via an ergodicity breaking non-equilibrium phase transition
\cite{Julicher95}. While a proof of principle for this so-called
spontaneous ratchet effect has been provided theoretically with
several different ``minimal models'' \cite{Julicher95,Reimann02},
we explore here for the first time a \emph{real} system, namely a
vibrofluidized granular gas, experimentally, numerically, and
analytically.

Transport in granular systems due to a ratchet effect has been
previously demonstrated both for asymmetric potentials
\cite{Farkas99} and for dynamical asymmetries \cite{Krulle}. While
in both these cases even a single granular particle moves in a
preferential direction, the spontaneous ratchet effect we consider
here is a genuine {\em collective phenomenon}. The underlying
mechanism is the enhanced energy loss in already dense regions of
a granular gas due to the inelastic collisions, resulting in a
spontaneous separation into dense and dilute regions
\cite{Goldhirsch-Jaeger-Kadanoff}. This clustering effect can be
controlled by confining the particles to a series of compartments
that are connected by small apertures (slits) and subjected to
gravity \cite{Eggers,Brey}. When the granular system is fluidized
by shaking it vertically, the particles cluster into the
compartments which are initially or due to random fluctuations
slightly denser, until a dynamical equilibrium is established in
which the particle flux out of a compartment is balanced by the
influx coming from its neighbors.

Hence, all ingredients for a spontaneous ratchet effect are
present: First, random fluctuations in the form of deterministic
noise are created intrinsically by the chaotic dynamics. Second,
the system is kept away from thermal equilibrium by the vertical
vibrations. Third, in a perfectly symmetric, periodic
compartmentalization, stable steady states with a spontaneously
broken symmetry yet still respecting the periodicity seem
possible, but are not at all obvious: their actual existence is
the main finding of our present work.

\begin{figure}
\begin{center}
\includegraphics*[scale=0.28]{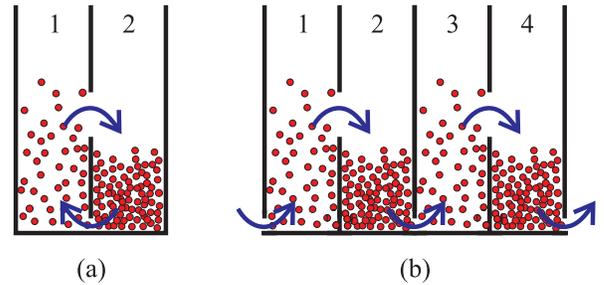}
\caption[]{\small (a) Schematic cross section of a granular
fountain exhibiting spontaneous symmetry breaking into a ``cold''
and a ``hot'' compartment and a concomitant spontaneous circular
particle flow. (b) By folding out the geometry of several adjacent
fountains, and adding cyclical boundary conditions, a granular
ratchet is obtained with almost unchanged populations of the
respective compartments and the particle currents between them
(here shown for $\KK=4$ compartments).} \label{Fig1}
\end{center}
\end{figure}

We start with a granular gas in a container that is divided into
$\KK =2$ equal compartments by a wall with a slit at a certain
height $h$. At vigorous shaking, this system has a unique steady
state with an equal population of both compartments \cite{Eggers}.
Upon reducing the driving, this symmetric state loses stability
via a continuous phase transition, resulting in a pair of stable
solutions with spontaneously broken symmetry: They are clustered
states with one diluted and one crowded compartment \cite{Eggers}.
If we now add a sufficiently small hole at the bottom of the wall
separating the compartments, the clustered states are expected to
subsist (see Fig.~\ref{Fig1}a), resulting in a particle flow
through the hole, balanced by a flow through the slit in the
opposite direction. We thus expect a \emph{``granular fountain''},
i.e. a spontaneous convective flow, imposed upon the system by the
clustering phenomenon.

\begin{figure}
\begin{center}
\includegraphics*[scale=0.4]{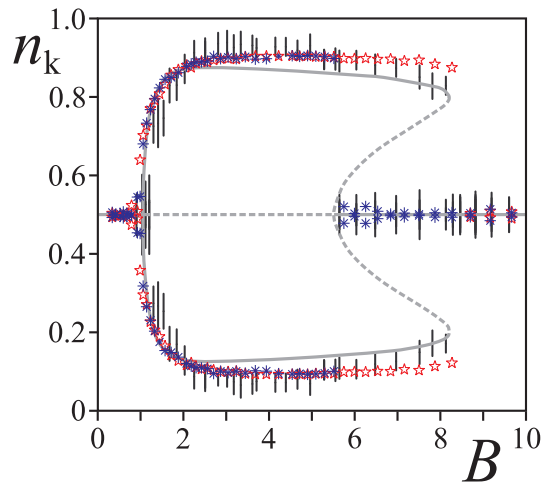}
\caption[]{\small Bifurcation diagram for the granular fountain.
Depicted are the steady-state fractions of particles $n_k$, $k\in
\{1,2\}$, in the two compartments from Fig.~\ref{Fig1}a for
different values of the driving frequency $f$ in units of $\BB
\propto f^{-2}$ from Eq. (\ref{eq1}). Experimental results are
indicated by the black error bars; the other symbols represent
molecular dynamics simulations using uniform (blue asterisks) and
clustered (red stars) initial conditions. The grey lines represent
the theoretical prediction from the flux model
(\ref{eq1})-(\ref{eq3}) for $\lambda = 0.018$: solid lines
correspond to stable configurations; dashed lines to unstable
ones.} \label{Fig2}
\end{center}
\end{figure}

To verify these predictions, we conducted both experiments and
molecular dynamics simulations. In the {\em experiments} we used
$\NN =407$ beads of stainless steel (radius $r = 1.18$ mm, normal
restitution coefficient $\ee \approx 0.9$) in a high perspex
container with a ground surface of 41.5 $\times$ $25.0$ mm$^2$.
When the particles are at rest, this corresponds to a filling
level of 2.0 layers. The box was divided into $\KK =2$ equal
compartments by a wall, with a horizontal slit of 25.0 mm by 5.0
mm, starting at height $h = 25.0$ mm, and a hole of dimensions
$3.0$ mm $\times$ $3.0$ mm at the bottom. The entire device was
subjected to vertical, sinusoidal vibrations with variable
frequency $f$ (ranging from 47 Hz to 150 Hz) and fixed amplitude
$a = 1.0$ mm. The results are given in Fig.~\ref{Fig2} in terms of
the dimensionless control parameter
\begin{equation}
\BB = 4 \pi \, \frac{gh}{a^2 f^2}(1-\ee^2)^2 \, \left( \frac{r^2
\NN}{\Omega \KK} \right)^2 , \label{eq1}
\end{equation}
which is motivated \cite{Eggers} from the kinetic theory for
dilute granular gases (see also Eq. (\ref{eq2}) below). Here
$\Omega$ ($= 19.7 \times 25.0$ mm$^2$) is the surface area of each
compartment, and $g=9.81$ m/s$^2$. As anticipated, \emph{vigorous
shaking} (small $\BB$) leads to a symmetric steady state, giving
way to a clustered state with the predicted fountain effect for
\emph{intermediate driving}. For very \emph{weak driving} (large
$\BB$) the system returns to equipartition via the hole at the
bottom of the container, since the flux through the slit (cf.
Fig.~\ref{Fig1}a) becomes negligibly small. Furthermore, a
pronounced hysteresis is observed in Fig.~\ref{Fig2} for
$\BB$-values between $5.5$ and $8.2$.

Fig.~\ref{Fig2} also contains the results of molecular dynamics
{\em simulations} in a setup comparable to the experiment: $K=2$
rectangular compartments with ground area $19.4 \times 25.0$
mm$^2$ and filled with $\NN =400$ particles of radius $r =1.18$ mm
and normal restitution coefficient $\ee = 0.9$ \cite{footnote0}.
While the slit at height $h$ was as in the experiment, the hole at
the bottom was chosen slightly larger ($4.2 \times 4.2$ mm$^2$) to
compensate for the vanishing thickness of the apertures and the
neglected dissipation from particle-wall collisions in the
simulations. Experiments and simulations are seen to agree well.

These findings can be explained in the
spirit of a previously established {\em flux model} \cite{Eggers} with
the help of two flux functions $F$ and $G$ describing the number
of particles which escape out of a compartment per time unit
through the slit and the hole, respectively:
\begin{eqnarray}\label{eq2}
\nonumber F(n_k) &=& \AAA\, n_k^2\, e^{-\BB \KK^2 n_k^2} ,\\
G(n_k) &=& \AAA\, \lambda\, n_k^2 .
\end{eqnarray}
Here, $\BB$ is defined in (\ref{eq1}), $n_k$ is the fraction of
particles in compartment $k\in\{1, ... ,\KK\}$ (hence
$n_k\in[0,1]$, $\Sigma_1^K n_k=1$), and $\AAA$ essentially
accounts for the size of the slit, $[A]=$ s$^{-1}$. The flux
$F(n_k)$ grows from zero for $n_k=0$ to a maximum at
$n_k=(\BB\KK^2)^{-1/2}$, and then decreases asymptotically towards
zero as a result of the inelastic collisions between the
particles. The flux through the hole, $G(n_k)$, is obtained by
taking the limit $h \to 0$ of $F(n_k)$. The ratio $\lambda$ of
$G(n_k)$ and $F(n_k)$ in the low density limit is determined by
the ratio of the surface areas of the hole and the slit, and also
accounts for the anisotropy of the velocity distribution just
above the bottom. Obviously, $\lambda$ should be chosen
considerably smaller than unity, otherwise the flux through the
hole will completely overpower that through the slit, and hence
the clustering effect. For our experimental setup one finds as a
rough estimate $\lambda \approx 0.02$ under the assumption that
the finite particle radius effectively reduces the dimensions of
slit and hole by $1.8$ mm and neglecting any velocity
anisotropies.

In the fountain geometry ($K=2$) from Fig.~\ref{Fig1}a the
time-evolution of the relative populations $n_1=n_1(t)$ and
$n_2=1-n_1$ follows by adding up all the fluxes through slit and
hole:
\begin{equation}\label{eq3}
    \dot{n}_1 = F(1-n_1) + G(1-n_1) - F(n_1) - G(n_1)
\end{equation}
The uniform distribution $n_1=n_2=1/2$ is a stationary solution
for all $\BB$, but is linearly stable only if the Jacobian
$-2F'(1/2)-2G'(1/2) = -2\AAA [(1-\BB)\exp(-\BB) + \lambda]$ is
negative. This is the case when $\BB$ is either small or large. If
$\lambda < e^{-2}$ there exists a $\BB$-interval for which the
uniform distribution is unstable, and one can show that now the
only stable solutions of (\ref{eq3}) are a symmetric pair of
clustered, fountain states. When fitting with $\lambda=0.018$, the
comparison of the predictions of the flux model with the
experiment in Fig.~\ref{Fig2} is fairly good; especially the
bifurcation and the hysteresis are well reproduced.

Convection rolls somewhat similar to our present granular fountain
but governed by different physical mechanisms are well known for
ordinary liquids and gases out of equilibrium and also for
granular systems at very high densities. For a dilute granular
gas, as we consider it here, convection in a single container has
been reported recently \cite{Ramirez00}, but neither a
compartmentalized setup nor a theoretical modeling have been
addressed before.

We now proceed towards the \emph{``granular ratchet''} setup in
Fig.~\ref{Fig1}b: Starting with any number of copies of the
granular fountain placed next to each other and closed cyclically,
we then move each hole to its adjacent wall. Within the flux model
(\ref{eq1}) - (\ref{eq2}) this amounts to replacing (\ref{eq3}) by
\begin{eqnarray}\label{eq4}
\nonumber
\!\!\!\!\!\!\dot{n}_{k} & \!\!\! = \!\!\! & G(n_{k+1}) + F(n_{k-1}) - F(n_{k}) -
G(n_{k}), \,\, \textrm{even $k$,}\\
\!\!\!\!\!\!\dot{n}_{k} & \!\!\! = \!\!\! & F(n_{k+1}) + G(n_{k-1}) - F(n_{k}) - G(n_{k}), \,\,
\textrm{odd $k$}
\end{eqnarray}
with an even number of compartments $\KK$ and periodic boundary
conditions $n_{\KK+k}=n_k$. Due to (\ref{eq2}) it follows that any
periodically continued steady state solution of the original
fountain geometry translates into a steady state solution of the
extended granular ratchet geometry; the particle currents through
the slits remain unchanged, and those through the holes are simply
inverted. Less obvious is the stability of these solutions and the
possible coexistence of further stable solutions:

\begin{figure}
\begin{center}
\includegraphics*[scale=0.4]{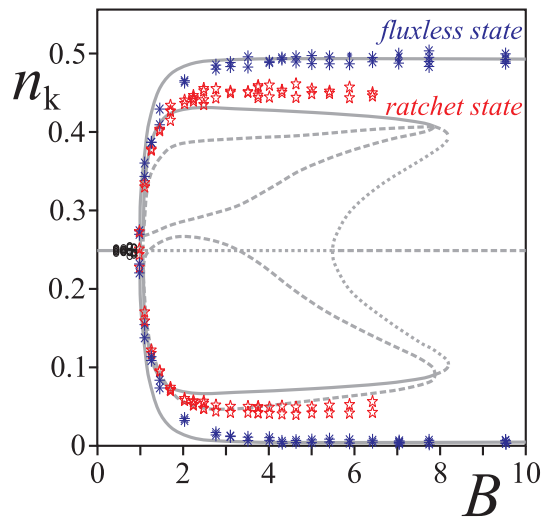}
\caption[]{\small Bifurcation diagram of the $\KK =4$ granular
ratchet. Symbols: results from molecular dynamics simulations.
Grey lines: prediction from the flux model (1), (2), (4) for
$\lambda = 0.018$. Solid lines correspond to stable
configurations; dashed lines to unstable ones. For further details
regarding parameters and methods see main text. For small $\BB$
the uniform distribution, $n_1=n_2=n_3=n_4=1/4$, is the unique
steady state. At $\BB=1$ a pair of symmetry broken stable states
takes over, namely $n_2=n_3\not = n_4=n_1$, indicated by blue
asterisks. They exhibit no net particle flow and are called
\emph{fluxless clustered states}. For slightly larger $\BB$
another pair of stable states emerges, both of the form
$n_1=n_3\not = n_2=n_4$ (see Fig.~\ref{Fig1}b), indicated by red
stars. They do carry a net flow (cf. Fig.~\ref{Fig4}) and are
called \emph{ratchet states}. For $\BB>6.8$ only the fluxless
stable states survive.} \label{Fig3}
\end{center}
\end{figure}

\begin{figure}
\begin{center}
\includegraphics*[scale=0.9]{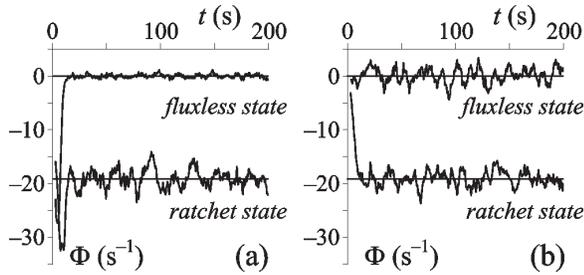}
\caption[]{\small Running time average (over 5 seconds) of the net
particle flux $\Phi (t)$ for the same molecular dynamics as in
Fig.~\ref{Fig3} with $\BB=3.0$, measured through the rightmost
slit (a), and through the rightmost hole (b) in Fig.~\ref{Fig1}b.
The upper (blue) curves in each plot correspond to the clustered
state with densely populated leftmost and rightmost compartments
in Fig.~\ref{Fig1}b. The lower (red) curves belong to the ratchet
state with densely populated compartments 1 and 3 in
Fig.~\ref{Fig1}b. For both states the straight line corresponds to
the long-time asymptote of the particle current.} \label{Fig4}
\end{center}
\end{figure}

In the simplest case of a granular ratchet consisting of just
$\KK=2$ periodically closed compartments, one obtains, as
expected, a bifurcation diagram (not shown) which practically
coincides with that in Fig.~\ref{Fig2}, and similarly for the
particle currents. Turning to $\KK=4$ compartments as depicted in
Fig.~\ref{Fig1}b, we have performed molecular dynamics simulations
with the same dimensions of compartment, hole, and slit and the
same average particle number $\NN / \KK = 200$ per compartment as
in the $\KK =2$ fountain case. The resulting bifurcation diagram
in Fig.~\ref{Fig3} displays as its most prominent new feature
\emph{two different types of coexisting stable states} with
spontaneously broken symmetry for not too large $\BB$-values: (i)
\emph{Ratchet states} with alternating dense and dilute
compartments and a resulting finite directed particle flux, as
quantitatively exemplified in Fig.~\ref{Fig4}: As the system
evolves towards its stable steady state, both fluxes in
Fig.~\ref{Fig4} converge to the same constant mean. Its finite
value is the most immediate signature of a spontaneous ratchet
effect. (ii) \emph{Fluxless clustered states} with alternating
pairs of dense and dilute compartments and no resulting net
particle flux, see Fig.~\ref{Fig4}. Our experimental data (not
shown) on the $\KK=4$ ratchet are in agreement with these
findings.

The intricate bifurcation diagrams for even larger $\KK$ will be
presented elsewhere. The main (and quite plausible) finding is
that current carrying ratchet states with alternating dense and
dilute compartments are always stable solutions within an entire
interval of $\BB$-values. Moreover, one finds an increasing number
of coexisting states with spontaneously broken symmetry. These are
long-lived hybrid states, containing both ratchet-like regions
(with alternating dense and dilute compartments) and fluxless
clustered regions.

\begin{figure}
\begin{center}
\includegraphics*[scale=0.32]{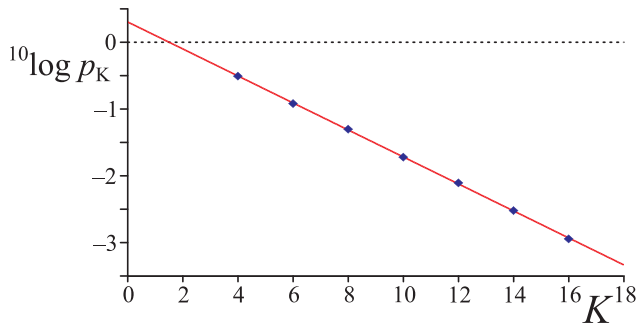}
\caption[]{\small Fraction $p_\KK$ of ratchet solutions versus
number $\KK (\geq 4)$ of compartments in a granular ratchet system
with a constant initial particle number $N/\KK$ per compartment.
Symbols represent numerical solutions of the (deterministic) flux
model (1), (2), (4) for $\BB=3$ and $\lambda=0.018$, averaged over
initial distributions with a superimposed randomization, mimicking
fluctuations due to the finite particle number within each
compartment. The straight line is a linear best fit.} \label{Fig5}
\end{center}
\end{figure}

Which specific steady state solution the system will settle in
strongly depends on the initial condition, and at phase boundaries
also on random fluctuations. An interesting example of the latter
type is a uniform initial distribution. The corresponding
probability $p_\KK$ of ending up in a ratchet state is depicted in
Fig.~\ref{Fig5}. For an intuitive explanation we view the system
as $\KK /2$ identical ``dimers'', each consisting of two
compartments with a common slit at height $h$, and connected with
each other by holes. Since the holes which connect neighboring
dimers are much smaller than the slits, inside each dimer a
quasi-steady distribution is approached much faster than the full
equilibration between all dimers. Hence, an initially uniform
distribution evolves inside each dimer towards either of two
clustered states with probability $q=1/2$, independently of the
other dimers. This implies that a granular ratchet system with
$\KK$ compartments will end in one of the two ratchet states with
alternating dense and dilute compartments with probability
\begin{equation}\label{eq5}
    p_\KK = 2\, q^{\KK /2} \ .
\end{equation}
This prediction is in excellent agreement with the linear best fit
$p_\KK=2.02\,(0.394)^{\KK/2}$ in Fig.~\ref{Fig5}. The reduced
$q$-value can be attributed to the fact that the dimers are
actually not entirely independent due the finite hole size.

In practice, a ratchet state spontaneously arises only with a
small probability for large $\KK$, as far as uniform initial
conditions are concerned. This situation is comparable to other
systems with spontaneous symmetry breaking, e.g. a ferromagnet
\cite{footnote1}: Just like a piece of iron picked from the shelf
normally appears to be ``unmagnetized'' and an external magnetic
field is needed to produce a magnetization on a global scale, a
ratchet state only emerges with appreciable probability for large
$\KK$ if there is some initial bias towards this state. Without
breaking the symmetry of the setup, this could be realized (in a
reproducible direction, if so desired) by applying a periodically
modulated initial particle distribution or a small external force
in the horizontal direction in Fig.~\ref{Fig1}b, acting during a
certain preparatory time span.

In conclusion, we have exploited the clustering effect to create
spontaneous directed transport in symmetrically compartmentalized
granular gases. In the granular fountain it takes the shape of a
convection roll, and in the granular ratchet it appears as a
current perpendicular to the vertical driving. In both cases, the
directed transport arises as a collective effect of the
stochastically colliding particles.

\acknowledgments{We thank Christof Kr\"ulle for pointing out to us
Refs. \cite{Krulle}.
This work is part of the research program of the
Stichting FOM, which is financially supported by NWO. P.R. was
supported by the Deutsche Forschungsgemeinschaft under SFB 613
(Teilprojekt K7) and HA-1517/13-4. }


\begin{thebibliography}{99}
\bibitem{footnote0}
The simulation uses a hard-sphere collision model in which the
dissipation is described by 3 parameters: The coefficients of
normal restitution, tangential restitution and friction. The
latter two were set equal to their ideal (dissipationless) values.
\bibitem{footnote1} For the analogy, the dimers of the ratchet system
should be compared to the single ferromagnetic domains in the
piece of ferromagnetic material.
\end{thebibliography}
\end{document}